\documentclass[showpacs,preprintnumbers,amsmath,amssymb,prb,twocolumn]{revtex4}

\def\XXint#1#2#3{{\setbox0=\hbox{$#1{#2#3}{\int}$}
     \vcenter{\hbox{$#2#3$}}\kern-.5\wd0}}

\usepackage{graphicx}
\usepackage{subfigure}
\usepackage{bm}

\begin{document}

\title{Spiral plane flops in frustrated helimagnets in external magnetic field}

\author{O.\ I.\ Utesov$^{1,2}$}
\email{utiosov@gmail.com}
\author{A.\ V.\ Syromyatnikov$^{1,3}$}
\email{asyromyatnikov@yandex.ru}

\affiliation{$^1$National Research Center ``Kurchatov Institute'' B.P.\ Konstantinov Petersburg Nuclear Physics Institute, Gatchina 188300, Russia}
\affiliation{$^2$St. Petersburg Academic University - Nanotechnology Research and Education Centre of the Russian Academy of Sciences, 194021 St.\ Petersburg, Russia}
\affiliation{$^3$St.\ Petersburg State University, 7/9 Universitetskaya nab., St.\ Petersburg, 199034
Russia}

\date{\today}

\begin{abstract}

We discuss theoretically frustrated Heisenberg spiral magnets in magnetic field $\bf H$. We demonstrate that small anisotropic spin interactions (single-ion biaxial anisotropy or dipolar forces) select the plane in which spins rotate (spiral plane) and can lead to the spiral plane flop upon in-plane field increasing. Expressions for the critical fields $H_{flop}$ are derived. It is shown that measuring of $H_{flop}$ is an efficient and simple method of quantifying the anisotropy in the system (as the measurement of spin-flop fields in collinear magnets with axial anisotropy). Corresponding recent experiments are considered in spiral magnets some of which are multiferroics of spin origin.

\end{abstract}


\maketitle

\section{Introduction}
\label{Intro}

%Frustration usually have a dramatic influence on properties of magnetic systems. It leads to new phenomena which have attracted significant interest recently, e.g. various spin-liquid phases, novel phase transitions, and order-by-disorder phenomena.~\cite{balents}
%For example, frustration changes the type of transitions to magnetically ordered phases in frustrated Heisenberg magnets with a spiral magnetic ordering. The type of the phase transition changes because the order parameter acquires additional symmetry elements. In three-dimensional (3D) Heisenberg helimagnets instead of second order phase transition in non-frustrated systems there is a first-order one in frustrated magnets; in 3D $XY$ systems novel pseudo-universal behavior occurs. \cite{kawa} In two-dimensional $XY$ systems frustration leads to the stabilization of a chiral spin-liquid phase upon cooling before the onset of Berezinskii-Kosterlitz-Thouless transition. \cite{sasha}

Multiferroics with coexisting magnetic and ferroelectric orders have attracted a lot of attention recently. \cite{nagaosa} The possibility to realize cross-control between magnetism and electricity in such compounds would lead to many desirable applications. For instance, strong enough magnetoelectric coupling would allow to manage magnetic memory by electric field. \cite{Khomskii} In so-called multiferroics of spin origin ferroelectricity is induced by some types of magnetic ordering and magnetoelectric coupling in such compounds is discovered to be strong. \cite{nagaosa,Cheong2007,Tokura2009}. There are three main mechanisms of ferroelectricity of spin origin: exchange-striction mechanism, inverse Dzyaloshinskii-Moriya (DM) mechanism and spin-dependent $p-d$ hybridization mechanism. \cite{nagaosa} Non-collinear spin ordering induced, e.g., by frustration is indispensable for the second and the third mechanisms.
%Therefore, frustration plays important role in many multiferroics of spin origin producing short period spiral.

While appearance of non-collinear magnetic textures in frustrated helimagnets is mainly caused by the competition between different exchange couplings, fine details of the spin ordering depend on usually weak low-symmetry relativistic interactions (anisotropy and dipolar forces). In particular, they fix the plane in which spins rotate (spiral plane) and, in turn, the direction of the electric polarization $\bf P$ which is related with the spiral plane orientation. \cite{nagaosa} The smallness of the anisotropic interactions opens a way to handle orientation of the spiral plane and $\bf P$ by, e.g., small magnetic field. \cite{Tokura2009}

It is well known that in collinear antiferromagnet a spin-flop transition of the first-order type takes place in magnetic field $\bf H$ applied along easy axis. \cite{Neel1936,white} Sublattices magnetizations stay parallel to $\bf H$ at $H<H_{flop}$ and they become nearly perpendicular to the field after the flop at $H>H_{flop}$ forming a canted antiferromagnetic spin arrangement. Well known relations are $H_{flop}\sim S\sqrt{DJ}\ll H_s\sim SJ$, where $S$ is the spin value, $D\ll J$ is the anisotropy value, $J$ is the exchange coupling constant, and $H_s$ is the saturation field. \cite{Neel1936,white}

\begin{figure}
  \centering
  \includegraphics[width=6cm]{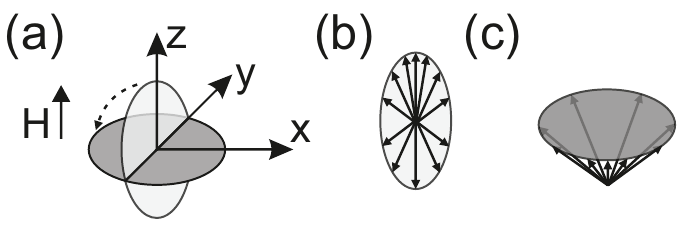}\\
  \caption{(a) Illustration of the flop of the plane in which spins rotate (spiral plane) upon in-plane field $\bf H$ increasing. The spiral plane containing the easy $z$ and the middle $y$ axes flops to $xy$ plane. (b) Spins lie in $yz$ plane at small field and form a helix slightly deformed by the field and anisotropy. (c) Spin arrangement after the flop (conical spiral).
	 \label{Fig1}}
\end{figure}

A similar phenomenon has been observed both experimentally (see, e.g., Refs.~\cite{Taniguchi2006,Schrettle2008,Banks2009},) and numerically (see, e.g., Refs.~\cite{Yan2014,zh}) in frustrated Heisenberg spiral magnets. Without anisotropy, the spiral plane is perpendicular to any finite $\bf H$. On the other hand, the spiral plane can be fixed by anisotropic interactions so that the spiral order is only slightly deformed by small in-plane magnetic field. However the spiral plane flops at some critical field $H_{flop}$ and becomes perpendicular to $\bf H$ at $H>H_{flop}$ as it is illustrated by Fig.~\ref{Fig1}. To the best of our knowledge, the spiral plane flops have not been described analytically so far. It is the aim of the present paper to fill up this gap.
% (c.f. chiral soliton lattice (CSL) in long-period spiral magnets, see Refs.~\cite{Dzyal1965, Izyumov1984}).
%Competing spin texture is the conical helix with the plane perpendicular to the external field. It loses energy in external magnetic field faster than the first one (it is proved for frustrated short-period spirals below). Thus, when magnetic field is large enough to overcome difference in the anisotropy energy the conical spiral becomes preferable and the spiral plane flop takes place. This phenomenon is illustrated in Fig.~\ref{Fig1}. If the anisotropy is small in comparison with the exchange interaction the flop field is much smaller than the saturation field.

In Sec.~\ref{Theor1} we discuss in details a simple model of frustrated Heisenberg magnet with small single-ion biaxial anisotropy. At $H=0$, a slightly distorted spiral ordering arises in the classical ground state of the system, where spins rotate in the plane containing easy and middle axes. Spin arrangement and expressions for $H_{flop}$ are found analytically for field directed along principal axes. We show that similar to collinear magnets $H_{flop}\sim S\sqrt{DJ}\ll H_s\sim SJ$.

It is well known that dipolar forces can be the main anisotropic interaction in helimagnets containing magnetic ions with $L=0$ (e.g., Mn$^{2+}$ and Eu$^{2+}$) in which anisotropy of spin-orbit origin is strongly suppressed. In particular, magneto-dipolar interaction was shown to be important for the description of transitions in many multiferroics of spin origin. \cite{Utesov2017} Then, we discuss in Sec.~\ref{Theor2} the spiral plane flops in frustrated Heisenberg magnets with dipolar interaction. The results obtained are qualitatively similar to those observed for the system with biaxial anisotropy.

%Then, we turn to frustrated helimagnets with the dipolar forces (which are always present in real materials). Their impact on the temperature phase transitions in such systems was discussed recently in Ref.~\cite{Utesov2017}. Magneto-dipolar interaction was shown to be important for the description of the transitions in many multiferroics of spin origin~\cite{Utesov2017}. We show, that with the same accuracy of calculations as in the case of biaxial anisotropy, the dipolar forces lead to the same results. Instead of the anisotropy constants dipolar tensor provides three different eigenvalues and corresponding easy, medium and hard directions in the spiral vector point. They determine the behaviour of the system in the external magnetic field and altogether with the exchange interaction determine the spiral plane flop field.

In Sec.~\ref{secarb} we analyze the systems in arbitrary directed magnetic field. We find that spiral plane flops can happen only if the external magnetic field lies in the spiral plane stabilized at $H=0$. In contrast to collinear magnets, where the spin flop takes place only for a very narrow range of field directions along the easy axis, \cite{bogdan} the spiral plane flop occurs for any field direction within the plane. Corresponding expressions for $H_{flop}$ are derived.

In Sec.~\ref{appsec} using our theory we describe experimentally observed field-induced spiral plane reorientations in various compounds including multiferroics of spin origin. We believe that our results could be useful for interpreting experimental data in many frustrated helimagnets in external magnetic field.

We point out in conclusion (Sec.~\ref{conc}) that measurement of $H_{flop}$ provides an easy and efficient way to quantify the anisotropy in frustrated helical magnets.

\section{Spiral plane flop in frustrated helimagnet with biaxial anisotropy}
\label{Theor1}

In this section we consider a simple model containing frustrating exchange interaction and a small single-ion biaxial anisotropy. We assume that the frustration leads to a spiral in the classical ground state.

\subsection{Model and general consideration}

The system Hamiltonian reads as
\begin{eqnarray}
 \label{ham1}
  \mathcal{H} &=& \mathcal{H}_{ex} + \mathcal{H}_{an} + \mathcal{H}_{z}, \nonumber \\
  \mathcal{H}_{ex} &=& -\frac12 \sum_{i,j} J_{ij} \left(\mathbf{S}_i \cdot \mathbf{S}_j\right), \\
  \mathcal{H}_{an} &=& - \sum_i \left[ D(S_i^z)^2 + E (S_i^y)^2\right], \nonumber \\
  \mathcal{H}_z &=& - \sum_i \left(\mathbf{h} \cdot \mathbf{S}_i\right),\nonumber
\end{eqnarray}
where $D > E > 0$, ${\bf h}=g \mu_B {\bf H}$, $x$ and $z$ axes are the hard and the easy ones, respectively, there is one spin in a unit cell, and the lattice is assumed arbitrary in all general derivations below. After the Fourier transform
\begin{equation}
\label{four1}
  \mathbf{S}_j = \frac{1}{\sqrt{N}} \sum_\mathbf{q} \mathbf{S}_\mathbf{q} e^{i \mathbf{q} \mathbf{R}_j},
\end{equation}
where $N$ is the number of spins in the lattice, Hamiltonian \eqref{ham1} acquires the following form:
\begin{eqnarray}
  \label{ex2}
  \mathcal{H}_{ex} &=& -\frac12 \sum_\mathbf{q} J_\mathbf{q} \left(\mathbf{S}_\mathbf{q} \cdot \mathbf{S}_{-\mathbf{q}}\right), \\
	\label{an21}
  \mathcal{H}_{an} &=& - \sum_\mathbf{q}\left[ D S^z_\mathbf{q} S^z_{-\mathbf{q}} + E S^y_\mathbf{q} S^y_{-\mathbf{q}}\right], \\
	\label{z21}
 \mathcal{H}_z &=& - \sqrt{N} \left(\mathbf{h} \cdot \mathbf{S}_{\bf 0}\right).
\end{eqnarray}

We assume that $J_\mathbf{q}$ has two equivalent maxima at ${\bf q}=\pm {\mathbf{k}}$ so that a plane spiral arises in the classical ground state at $h=D=E=0$. The plane at which spins lie can be fixed by small anisotropy and/or magnetic field which can also distort the spiral order.

For theoretical description of a cone helix, we introduce the local right-hand orthogonal coordinate frame at $j$-th site (see Ref.~\cite{Maleyev2006})
\begin{eqnarray}
  \label{bas1}
  \hat{\zeta}_j &=& (\hat{a} \cos{\mathbf{k}\mathbf{R}_j} + \hat{b} \sin{\mathbf{k}\mathbf{R}_j}) \cos{\alpha} + \hat{c} \sin{\alpha},\nonumber \\
	\hat{\eta}_j &=&  - \hat{a} \sin{\mathbf{k}\mathbf{R}_j} + \hat{b} \cos{\mathbf{k}\mathbf{R}_j}, \\
 \hat{\xi}_j &=& - (\hat{a} \cos{\mathbf{k}\mathbf{R}_j} + \hat{b} \sin{\mathbf{k}\mathbf{R}_j}) \sin{\alpha} + \hat{c} \cos{\alpha},\nonumber
\end{eqnarray}
where $\hat{a}$, $\hat{b}$, and $\hat{c}$ are some mutually orthogonal unit vectors, and $\alpha$ is a cone angle ($\alpha=0$ in the plane spiral). Then, the spin at $j$-th site is expressed as
\begin{equation}
\label{spin1}
  \mathbf{S}_j = S_j^\zeta \hat{\zeta}_j + S_j^\eta \hat{\eta}_j + S_j^\xi \hat{\xi}_j,
\end{equation}
where
\begin{eqnarray}
\label{spinrep1}
  S^\zeta_j &=& S - a^\dagger_j a_j, \nonumber \\
  S^\eta_j &\simeq& \sqrt{\frac{S}{2}} \left( a_j + a^\dagger_j \right), \\
  S^\xi_j &\simeq& i \sqrt{\frac{S}{2}} \left( a^\dagger_j - a_j \right),\nonumber
\end{eqnarray}
is the Holstein-Primakoff transformation~\cite{Holstein1940} in which square roots a replaced by unity. It is convenient to rewrite local basis vectors \eqref{bas1} as
\begin{eqnarray}
  \label{bas2}
 \hat{\zeta}_j &=& (\mathbf{A} e^{i \mathbf{k} \mathbf{R}_j} + \mathbf{A}^* e^{- i \mathbf{k} \mathbf{R}_j}) \cos{\alpha} + \hat{c} \sin{\alpha}  \nonumber \\
 \hat{\eta}_j &=& i \mathbf{A} e^{i \mathbf{k} \mathbf{R}_j} - i \mathbf{A}^* e^{- i \mathbf{k} \mathbf{R}_j} \\
 \hat{\xi}_j &=& -(\mathbf{A} e^{i \mathbf{k} \mathbf{R}_j} + \mathbf{A}^* e^{- i \mathbf{k} \mathbf{R}_j}) \cos{\alpha} + \hat{c} \cos{\alpha},  \nonumber
\end{eqnarray}
where auxiliary vectors $\mathbf{A} = (\hat{a} - i\hat{b})/2$ and $\mathbf{A}^* = (\hat{a} + i\hat{b})/2$ are introduced. Then, we have from Eqs.~\eqref{spin1} and \eqref{bas2} after Fourier transform~\eqref{four1}
\begin{equation}
\label{spin2}
  \mathbf{S}_\mathbf{q} = S^A_\mathbf{q} \mathbf{A} + S^{A^*}_\mathbf{q} \mathbf{A}^* +S^c_\mathbf{q} \hat{c},
\end{equation}
where
\begin{eqnarray}
  \label{spin3}
 S^A_\mathbf{q} &=&  S^\zeta_{\mathbf{q} -\mathbf{k}} \cos{\alpha} + i S^\eta_{\mathbf{q}-\mathbf{k}} - S^\xi_{\mathbf{q}-\mathbf{k}} \sin{\alpha} ,  \nonumber\\
 S^{A^*}_\mathbf{q} &=& S^\zeta_{\mathbf{q}+\mathbf{k}} \cos{\alpha}  - i S^\eta_{\mathbf{q}+\mathbf{k}} - S^\xi_{\mathbf{q}+\mathbf{k}}  \sin{\alpha},\\
 S^c_\mathbf{q} &=&  S^\zeta_\mathbf{q} \sin{\alpha} + S^\xi_\mathbf{q} \cos{\alpha}.  \nonumber
\end{eqnarray}

Substituting Eqs.~\eqref{spin2} and \eqref{spin3} into Eqs.~\eqref{ex2} and \eqref{an21}, one obtains
\begin{eqnarray}
 \label{exch1}
  \nonumber \mathcal{H}_{ex} &=& -\frac{1}{2} \sum_\mathbf{q} \Bigl[ \left( \sin^2{\alpha} J_\mathbf{q} + \cos^2{\alpha} J_{\mathbf{q},\mathbf{k}} \right) S^\zeta_\mathbf{q} S^\zeta_\mathbf{-q}   \\
 \nonumber &+& J_{\mathbf{q},\mathbf{k}} S^\eta_\mathbf{q} S^\eta_\mathbf{-q} + \left( \cos^2{\alpha} J_\mathbf{q} + \sin^2{\alpha} J_{\mathbf{q},\mathbf{k}} \right) S^\xi_\mathbf{q} S^\xi_\mathbf{-q}  \\
  \nonumber &+& \sin{\alpha} \cos{\alpha} \left( J_\mathbf{q} - J_{\mathbf{q},\mathbf{k}} \right)  \left( S^\zeta_\mathbf{q} S^\xi_\mathbf{-q} + S^\xi_\mathbf{q} S^\zeta_\mathbf{-q} \right) \\
 \nonumber &+& i \cos{\alpha} N_{\mathbf{q},\mathbf{k}} \left( S^\eta_\mathbf{q} S^\zeta_\mathbf{-q} - S^\zeta_\mathbf{q} S^\eta_\mathbf{-q} \right) \\
 &+& i \sin{\alpha} N_{\mathbf{q},\mathbf{k}} \left( S^\xi_\mathbf{q} S^\eta_\mathbf{-q} - S^\eta_\mathbf{q} S^\xi_\mathbf{-q} \right) \Bigr],
\end{eqnarray}
where $J_{\mathbf{q},\mathbf{k}}=(J_{\mathbf{q}+\mathbf{k}}+J_{\mathbf{q}-\mathbf{k}})/2$ and $N_{\mathbf{q},\mathbf{k}}=(J_{\mathbf{q}+\mathbf{k}}-J_{\mathbf{q}-\mathbf{k}})/2$, and
\begin{eqnarray}
\label{an1}
 \mathcal{H}_{an} &=& -D \sum_\mathbf{q} \left( S^A_\mathbf{q} A_z + S^{A^*}_\mathbf{q} A^*_z + S^c_\mathbf{q} c_z \right)  \nonumber\\
  \nonumber  && \left( S^A_{-\mathbf{q}} A_z + S^{A^*}_{-\mathbf{q}} A^*_z + S^c_{-\mathbf{q}} c_z \right)  \\
 && - E \sum_\mathbf{q} \left( S^A_\mathbf{q} A_y + S^{A^*}_\mathbf{q} A^*_y + S^c_\mathbf{q} c_y \right) \nonumber\\
&& \left( S^A_{-\mathbf{q}} A_y + S^{A^*}_{-\mathbf{q}} A^*_y + S^c_{-\mathbf{q}} c_y \right).
\end{eqnarray}
%In general, these expressions are quite cumbersome. Then, we have to analyze them below only in some special cases.
%The form of Zeeman interaction is also dependent on the orientation of the spiral plane and the external magnetic field.

\subsection{Ground-state energy of the plane helix at finite anisotropy and $h=0$}
\label{YZnoField}

At zero field, the spin texture in the classical ground state is a slightly distorted (due to the anisotropy) spiral in which spins lie in $yz$ plane. Then, we take $\hat{a}=\mathbf{e}_y$, $\hat{b}=\mathbf{e}_z$, and $\hat{c}=\mathbf{e}_x$ in Eq.~\eqref{bas1}, where ${\bf e}_{x,y,z}$ are unit vectors directed along corresponding axes. To find the ground-state energy and the spin arrangement, we substitute Eqs.~\eqref{spinrep1} into Eqs.~\eqref{exch1} and \eqref{an1} and put $\alpha=0$. One obtains for the Hamiltonian as a result
%One can see that the second sum of Eq.~\eqref{an2} provides linear in bosonic operators terms (they arise from $S^\zeta_{0} S^\eta_{\pm 2 \mathbf{k}}$) and umklapp terms (they are bilinear in bosonic operators and violates the momentum conservation law by $\pm 2 \mathbf{k}$). Below we will neglect the umklapps, because for small anisotropy the main correction to the energy stems from the shift in operators due to the linear terms, and the umklapps lead to the second order corrections for the shifts. Thus, altogether with the exchange interaction~\eqref{exch1} (where we put $\alpha=0$) bosonic Hamiltonian obtains the following form:
\begin{eqnarray}
 \label{ham3}
\mathcal{H} &=& {\cal E}^{yz}_{0} + \mathcal{H}_1 + \mathcal{H}_2, \\
\label{eclass1}
  \frac1N {\cal E}^{yz}_{0} &=& - \frac{S^2 J_\mathbf{k}}{2} - \frac{S^2(D + E)}{2},\\
\label{h1}
 \frac{1}{\sqrt N} \mathcal{H}_{1an} &=& i (D-E) \left(\frac{S}{2}\right)^{3/2} \nonumber\\
&&\times\left( a_{-2\mathbf{k}} - a_{2\mathbf{k}} + a^\dagger_{2\mathbf{k}} - a^\dagger_{-2\mathbf{k}}\right), \\
\label{h2}
 \mathcal{H}_2 &=& \sum_\mathbf{q}
\left(C_\mathbf{q} a^\dagger_\mathbf{q} a_\mathbf{q} + B_\mathbf{q} \frac{a_\mathbf{q} a_{-\mathbf{q}} + a^\dagger _\mathbf{q} a^\dagger _{-\mathbf{q}}}{2}\right),
\end{eqnarray}
where
\begin{eqnarray}
  C_\mathbf{q} &=& \frac{S}{2} \left( 2J_\mathbf{k} - J_{\mathbf{q},\mathbf{k}} - J_\mathbf{q} + D + E \right), \\
  B_\mathbf{q} &=& -\frac{S}{2} \left( J_{\mathbf{q},\mathbf{k}} - J_\mathbf{q} + D + E \right).
\end{eqnarray}
We omit the so-called umklapp terms in $\mathcal{H}_2$ which have the form $a^\dagger_{\bf q}a_{{\bf q}\pm 2 \mathbf{k}}$, $a_{\bf q}a_{-{\bf q}\pm 2 \mathbf{k}}$, and $a^\dagger_{\bf q}a^\dagger_{-{\bf q}\pm 2 \mathbf{k}}$ and which are proportional to $D-E$. As it is explained below, their contribution to the ground-state energy and the spin arrangement is small.

Terms linear in Bose-operators $\mathcal{H}_{1an}$ arise in Hamiltonian \eqref{ham3} because we assume in derivation of Eqs.~\eqref{exch1} and \eqref{an1} that the spiral ordering is undisturbed (see Eq.~\eqref{bas1}).
To eliminate the linear terms from the Hamiltonian~\eqref{ham3}, we perform the following shift in operators:
\begin{eqnarray}
  \label{shift1}
   a_{2\mathbf{k}} &\mapsto& \rho_+ e^{i \varphi_+} + a_{2\mathbf{k}} ,
	\quad
	a^\dagger _{2\mathbf{k}} \mapsto \rho_+ e^{-i \varphi_+} + a^\dagger _{2\mathbf{k}}, \\
  a_{-2\mathbf{k}} &\mapsto& \rho_- e^{i \varphi_-} + a_{-2\mathbf{k}},
	\quad
	a^\dagger _{-2\mathbf{k}} \mapsto \rho_- e^{-i \varphi_-} + a^\dagger _{-2\mathbf{k}},\nonumber
\end{eqnarray}
where $\rho_\pm$ and $\varphi_\pm$ are real constants. Linear terms vanish if the following equalities hold:
\begin{eqnarray}
\label{eqna}
 && - i \frac{D-E}{2}S \sqrt{\frac{S}{2}} \sqrt{N} + C_{2\mathbf{k}} \rho_+ e^{-i \varphi_+} + B_{2\mathbf{k}} \rho_- e^{i \varphi_-} = 0,  \nonumber \\
  && i \frac{D-E}{2}S \sqrt{\frac{SN}{2}} + C_{2\mathbf{k}} \rho_- e^{-i \varphi_-} + B_{2\mathbf{k}} \rho_+ e^{i \varphi_+} = 0.
\end{eqnarray}
A solution of Eqs.~\eqref{eqna} has the form
\begin{eqnarray}
\label{shift3}
\varphi_+ &=& -\varphi_- = \pi/2,\nonumber\\
  \rho_+ &=& \rho_-
	= - \sqrt{N} \sqrt{\frac S2}\frac{D-E }{J_\mathbf{k}- J_{3 \mathbf{k} }}.
\end{eqnarray}
A correction $\Delta {\cal E}^{yz}_{an}$ to the constant ${\cal E}^{yz}_{0}$ also arises after shift \eqref{shift1} which has the form $-N(C_{2\bf k}\rho_+^2 + C_{-2\bf k}\rho_-^2 + (B_{2\bf k} + B_{-2\bf k})\rho_+\rho_-)/2$. Substituting Eqs.~\eqref{shift3} to this formula, one obtains
\begin{equation}
\label{encorr1}
  \frac1N \Delta {\cal E}^{yz}_{an} = -\frac{ S^2 (D-E)^2  }{2(J_\mathbf{k}- J_{3 \mathbf{k} })}.
\end{equation}

One has for the spin arrangement from Eqs.~\eqref{spinrep1}--\eqref{spin3} after taking into account shift \eqref{shift1} and Eqs.~\eqref{shift3}
\begin{eqnarray}
\label{arr}
  \mathbf{S}_j = && S \Biggl[ \mathbf{e}_z \left( 1 + \frac{D-E}{J_\mathbf{k}- J_{3 \mathbf{k} }} \right) \sin{\mathbf{k}\mathbf{R}_j}  \nonumber \\ &&+ \mathbf{e}_y \left( 1 - \frac{D-E}{J_\mathbf{k}- J_{3 \mathbf{k} }} \right) \cos{\mathbf{k}\mathbf{R}_j} \\ && + \frac{D-E}{J_\mathbf{k}- J_{3 \mathbf{k} }} \left( \mathbf{e}_z \sin{3 \mathbf{k}\mathbf{R}_j} + \mathbf{e}_y \cos{3 \mathbf{k}\mathbf{R}_j} \right) \Biggr]. \nonumber
\end{eqnarray}
Then, we obtain that the in-plane anisotropy leads to an elliptical distortion of the spiral and to the third harmonic of $\bf k$.

Umklapp terms would complicate considerably the above analysis. In particular, one would have to consider shifts of the form \eqref{shift1} for momenta $2n\bf k$, where $n$ is any integer. As a result, an infinite set of equations would arise instead of Eqs.~\eqref{eqna}. Fortunately, umklapp terms are proportional to $D-E$. Then, it is easy to realize that their contribution to Eqs.~\eqref{encorr1} and \eqref{arr} is of the third order in small parameter $(D-E)/J$ which can be safely neglected.

\subsection{Ground-state energy of the plane helix at finite anisotropy and in-plane magnetic field}

Let us take into account the in-plane magnetic field directed along $z$-axis. One obtains from Eqs.~\eqref{z21} and \eqref{spinrep1}--\eqref{spin3} the following contribution to ${\cal H}_1$:
\begin{equation}\label{zeem1}
   \frac{1}{\sqrt N} \mathcal{H}_{1z} = -\frac{ h}{2} \sqrt{\frac{S}{2}} \left(a_{\mathbf{k}} + a_{-\mathbf{k}} + a^\dagger _{\mathbf{k}} + a^\dagger _{-\mathbf{k}}\right)
\end{equation}
which contains Bose-operators on momenta $\pm\bf k$ rather than $\pm 2\bf k$ (cf.\ Eq.~\eqref{h1}).
To eliminate $\mathcal{H}_{1z}$, we perform a shift similar to Eq.~\eqref{shift1}
\begin{eqnarray}
  \label{shift4}
   a_{\mathbf{k}} &\mapsto& \tilde\rho_+ e^{i \tilde\varphi_+} + a_{\mathbf{k}} ,
	\quad
	a^\dagger _{\mathbf{k}} \mapsto \tilde\rho_+ e^{-i \tilde\varphi_+} + a^\dagger _{\mathbf{k}}, \\ \nonumber
  a_{-\mathbf{k}} &\mapsto& \tilde\rho_- e^{i \tilde\varphi_-} + a_{-\mathbf{k}},
	\quad
	a^\dagger _{-\mathbf{k}} \mapsto \tilde\rho_- e^{-i \tilde\varphi_-} + a^\dagger _{-\mathbf{k}}.
\end{eqnarray}
Linear terms vanish when
\begin{eqnarray}
  \nonumber &&-  \frac{h}{2} \sqrt{\frac{SN}{2}}  + C_{\mathbf{k}} \tilde\rho_+ e^{-i \tilde\varphi_+} + B_{\mathbf{k}} \tilde\rho_- e^{i \tilde\varphi_-} = 0, \\
  && - \frac{h}{2} \sqrt{\frac{SN}{2}} + C_{\mathbf{k}} \tilde\rho_- e^{-i \tilde\varphi_-} + B_{\mathbf{k}} \tilde\rho_+ e^{i \tilde\varphi_+} = 0
\end{eqnarray}
that gives
\begin{eqnarray}
\label{shift5}
\tilde\varphi_+ &=& \tilde\varphi_- = 0,\nonumber\\
  \tilde\rho_+ &=& \tilde\rho_- =  \frac{ h \sqrt{NS/2}}{S (2 J_\mathbf{k}- J_{\bf 0} -  J_{2 \mathbf{k}})}.
\end{eqnarray}
The correction to the ground-state energy appearing as a result of the shift \eqref{shift4} reads as
\begin{equation}
\label{encorr2}
  \frac1N\Delta {\cal E}^{yz}_z = -\frac{h^2   }{2(2 J_\mathbf{k}- J_{\bf 0} -  J_{2 \mathbf{k}})}.
\end{equation}
It can be shown that the correction to the ground-state energy for the field directed along $y$ axis is also given by Eq.~\eqref{encorr2}.

Thus, we obtain from Eqs.~\eqref{eclass1}, \eqref{encorr1}, and \eqref{encorr2} for the energy of the spiral in which all spins lie in $yz$ plane
\begin{eqnarray}
  \label{eYZ}
\frac1N  {\cal E}^{yz} &=& - \frac{S^2 J_\mathbf{k}}{2} - \frac{S^2(D + E)}{2} -\frac{ S^2 (D-E)^2  }{2(J_\mathbf{k}- J_{3 \mathbf{k} })}  \nonumber \\
 && -\frac{h^2}{2\left(2 J_\mathbf{k}- J_{\bf 0} -  J_{2 \mathbf{k}}\right)}.
\end{eqnarray}

\subsection{Ground-state energy of conical helix}

We calculate now the ground-state energy of the conical spiral in which spins rotate in $xy$ plane (see Fig.~\ref{Fig1}(c)). In this case, all spins are canted towards magnetic field direction (i.e., $z$-axis) and $\alpha\ne0$. It is convenient to take $\hat{a}=\mathbf{e}_x$, $\hat{b}=\mathbf{e}_y$, and $\hat{c}=\mathbf{e}_z$ in Eq.~\eqref{bas1}. The angle $\alpha$ is to be chosen to eliminate linear in $a_{\bf 0}$ and $a^\dagger_{\bf 0}$ terms in the Hamiltonian. As usual, these $\alpha$ values minimize the system classical energy having the form
\begin{eqnarray}
\label{eclass2}
\frac1N {\cal E}^{xy}_0 &=& -\frac{S^2 \left( J_{\bf 0} \sin^2{\alpha} + J_\mathbf{k} \cos^2{\alpha} \right)}{2} - S^2 D \sin^2{\alpha}  \nonumber \\
  && -\frac{S^2 E \cos^2{\alpha}}{2}- h S \sin{\alpha}.
\end{eqnarray}
The minimum of ${\cal E}^{xy}_0$ is achieved at
\begin{equation}
\label{alpha2}
  \sin{\alpha} = \frac{h}{S\left( J_\mathbf{k} - J_{\bf 0} -2D + E\right)}
	\approx \frac{h}{S\left( J_\mathbf{k} - J_{\bf 0}\right)}.
\end{equation}
We obtain from Eqs.~\eqref{eclass2} and \eqref{alpha2} in the leading orders in small parameters $E/J$, $D/J$, and $h/J$
\begin{equation}
\label{eclass3}
 \frac1N {\cal E}^{xy}_0 = - \frac{S^2 J_\mathbf{k}}{2} - \frac{S^2 E}{2} - \frac{h^2}{2 \left( J_\mathbf{k} - J_{\bf 0}\right) }.
\end{equation}
One has also to eliminate terms in the Hamiltonian linear in $a_{\pm2\bf k}$ and $a^\dagger_{\pm2\bf k}$ stemming from the anisotropy. Calculations similar to those performed above in Sec.~\ref{YZnoField} lead to the following correction to the ground-state energy (cf.\ Eq.~\eqref{encorr1}):
\begin{equation}\label{encorr3}
  \frac1N\Delta {\cal E}^{xy}_{an} = -\frac{ S^2 E^2}{2(J_\mathbf{k}- J_{3 \mathbf{k} })}.
\end{equation}

Thus, we obtain from Eqs.~\eqref{eclass3} and \eqref{encorr3} for the energy of the conical spiral in which spins rotate in $xy$-plane
\begin{eqnarray}
  \label{eXY}
 \frac1N {\cal E}^{xy} &=& - \frac{S^2 J_\mathbf{k}}{2} - \frac{S^2 E}{2}  -\frac{ S^2 E^2}{2(J_\mathbf{k}- J_{3 \mathbf{k} })}  \nonumber \\
 && - \frac{h^2}{2 \left( J_\mathbf{k} - J_{\bf 0}\right) }.
\end{eqnarray}

\subsection{Spiral plane flop in magnetic field}

Let us compare now energies ${\cal E}^{yz}$ and ${\cal E}^{xy}$ of the plane and the conical spirals given by Eqs.~\eqref{eYZ} and \eqref{eXY}, respectively. It is seen that ${\cal E}^{yz}<{\cal E}^{xy}$ at $h=0$. However, the field-correction in Eq.~\eqref{eXY} is smaller than that in Eq.~\eqref{eYZ} because $J_{\bf k}>J_{2\bf k}$ (remember, $J_{\bf q}$ is maximized at ${\bf q}=\pm{\bf k}$). Thus, ${\cal E}^{xy}$ becomes smaller than ${\cal E}^{yz}$ at $h>h_{flop}$, where $h_{flop}$ is determined in the leading order in small parameters by the equation
\begin{equation}
\label{hc11}
  S^2 D = \frac{h^2_{flop}}{ J_\mathbf{k} - J_{\bf 0}} - \frac{h^2_{flop}}{2 J_\mathbf{k}- J_{\bf 0} -  J_{2 \mathbf{k}}}.
\end{equation}
Then, the spiral plane flop takes place at the critical field $h_{flop}$ for which we have from Eq.~\eqref{hc11}
\begin{equation}
\label{hc12}
  h_{flop} = S \sqrt{D \tilde J},
\end{equation}
where
\begin{equation}
\label{jtild}
  \tilde{J} = \frac{\left( J_\mathbf{k} - J_{\bf 0}\right)\left(2 J_\mathbf{k}- J_{\bf 0} -  J_{2 \mathbf{k}}\right)}{J_\mathbf{k} - J_{2\mathbf{k}}} .
\end{equation}
%Evidently, that this magnetic field is much larger than the anisotropy, so before calculating further corrections with Eqs.~\eqref{eYZ} and~\eqref{eXY} one should take into account the umklapps neglected in Zeeman term~\eqref{zeem1}.
Notice that $h_{flop}\sim S\sqrt{DJ}$ is much smaller than the saturation field
\begin{equation}
\label{hs}
h_s = S \left( J_\mathbf{k}- J_{\bf 0}\right)
\end{equation}
 found from Eq.~\eqref{alpha2} because $h_s\sim SJ$.

The critical field $h_{flop}$ given by Eqs.~\eqref{hc12} and \eqref{jtild} is related to $h_s$ as
\begin{equation}
\label{hc13}
  h_{flop} = \sqrt{2 S D h_s}
\end{equation}
if the exchange interaction satisfies the condition $J_{\bf 0} \approx  J_{2 \mathbf{k}}$ in which case
\begin{equation}
\label{jta}
	\tilde{J} \approx 2\left( J_\mathbf{k}- J_{\bf 0}\right)=2h_s/S.
\end{equation}
One expects that the latter equality is fulfilled not so rare as soon as points ${\bf q}={\bf 0}$ and ${\bf q}={2\bf k}$ are symmetric according to ${\bf q}={\bf k}$ at which $J_{\bf q}$ is maximized. Eq.~\eqref{hc13} may be very useful in determination of the anisotropy value from experimentally obtained values of $h_{flop}$ and $h_s$. Interestingly, Eq.~\eqref{hc13} coincides with the spin-flop field in collinear magnets with small easy-axis anisotropy $D$.

As it follows from the above discussion, one should substitute $D$ by $E$ in Eqs.~\eqref{hc12} and \eqref{hc13} if the magnetic field is directed along $y$ axis.

\section{Spiral plane flop in frustrated helimagnet with dipolar forces}
\label{Theor2}

In this section, we show that small magneto-dipolar interaction has a similar impact on the spiral ordering as the biaxial anisotropy discussed above. The system Hamiltonian has the form \eqref{ham1}, where $\mathcal{H}_{an}$ should be replaced by
\begin{eqnarray}
 \label{ham4}
  \mathcal{H}_d &=& \frac12 \sum_{i,j} D^{\alpha \beta}_{ij} S^\alpha_i S^\beta_j, \\
	 {\cal D}^{\alpha \beta}_{ij} &=& \omega_0 \frac{v_0}{4 \pi} \left( \frac{1}{R_{ij}^3} - \frac{3 R_{ij}^\alpha R_{ij}^\beta }{R_{ij}^5}\right), \nonumber
\end{eqnarray}
where $v_0$ is the unit cell volume and
\begin{equation}\label{dipen}
  \omega_0 = 4 \pi \frac{(g \mu_B)^2}{v_0} \ll J
\end{equation}
is the characteristic dipolar energy. We have after Fourier transform~\eqref{four1}
\begin{equation}\label{dip2}
  \mathcal{H}_d = \frac12 \sum_\mathbf{q} {\cal D}^{\alpha \beta}_\mathbf{q} S^\alpha_\mathbf{q} S^\beta_{-\mathbf{q}}.
\end{equation}
Tensor ${\cal D}^{\alpha \beta}_\mathbf{q}/2$ has three eigenvalues $\lambda_1(\mathbf{q}) \geq \lambda_2(\mathbf{q}) \geq \lambda_3(\mathbf{q})$ corresponding to three orthogonal eigenvectors $\mathbf{v}_1(\mathbf{q})$, $\mathbf{v}_2(\mathbf{q})$, and $\mathbf{v}_3(\mathbf{q})$.

At $h=0$, the classical ground-state energy per spin $ - J_\mathbf{q} + ( \lambda_2(\mathbf{q}) + \lambda_3(\mathbf{q}))/2  $ is minimized at an incommensurate vector $\mathbf{k}$ which is close to the momentum maximizing $J_\mathbf{q}$. Then, $\mathbf{v}_1(\mathbf{k})$, $\mathbf{v}_2(\mathbf{k})$, and $\mathbf{v}_3(\mathbf{k})$ are the hard, the middle, and the easy axis for magnetization along which we direct $x$, $y$, and $z$ axes, respectively. Notice that $ {\cal D}^{\alpha \beta}_\mathbf{k}$ is diagonal in this basis. One obtains from Eqs.~\eqref{spinrep1}--\eqref{spin3} for terms linear in bosonic operators, which arise in Eq.~\eqref{dip2} only at $\mathbf{q}=\pm 2\mathbf{k}$
\begin{eqnarray}
\label{h1d}
	 \frac{1}{\sqrt N} {\cal H}_{1d} &=& i [ \lambda_2(\mathbf{k}) - \lambda_3(\mathbf{k})] \left(\frac{S}{2}\right)^{3/2} \nonumber\\
&&\times\left( a_{-2\mathbf{k}} - a_{2\mathbf{k}} + a^\dagger_{2\mathbf{k}} - a^\dagger_{-2\mathbf{k}}\right).
\end{eqnarray}
Linear terms \eqref{h1d} have the same form as those arisen in the case of biaxial anisotropy (see Eq.~\eqref{h1}). Corrections to the ground state energies can be calculated in much the same way as it is done above for the biaxial anisotropy.

As a result, one has to compare the following ground-state energies if the magnetic field is directed along $z$ axis (cf.\ Eqs.~\eqref{eYZ} and \eqref{eXY}):
\begin{eqnarray}
  \label{edipYZ}
  \frac1N{\cal E}^{yz} &=& - \frac{S^2 J_\mathbf{k}}{2} - \frac{S^2[2\lambda_1(\mathbf{k}) -  \lambda_2(\mathbf{k}) - \lambda_3(\mathbf{k})]}{2} \nonumber\\
	&& -\frac{h^2}{2\left(2 J_\mathbf{k}- J_{\bf 0} -  J_{2 \mathbf{k}}\right)}, \\
  \label{edipXY}
  \frac1N{\cal E}^{xy} &=& - \frac{S^2 J_\mathbf{k}}{2} - \frac{S^2 [\lambda_1(\mathbf{k})-\lambda_2(\mathbf{k})]}{2}
	\nonumber  \\ &&
	- \frac{h^2}{2 \left( J_\mathbf{k} - J_{\bf 0}\right) }.
\end{eqnarray}
The critical field value at which the spiral plane flop takes place reads as (cf.\ Eq.~\eqref{hc12})
\begin{equation}
\label{hc14}
  h_{flop} = S\sqrt{\left[\lambda_1(\mathbf{k})-\lambda_3(\mathbf{k})\right] \tilde J},
\end{equation}
where $\tilde{J}$ is given by Eq.~\eqref{jtild}. If the external magnetic field is along $y$ axis, the spiral plane flop occurs at
\begin{equation}
\label{hc15}
  h_{flop} = S\sqrt{\left[\lambda_1(\mathbf{k})-\lambda_2(\mathbf{k})\right]  \tilde J}.
\end{equation}
Eqs.~\eqref{hc14} and \eqref{hc15} can be related to $h_s$ using Eq.~\eqref{jta} if $J_{\bf 0} \approx  J_{2 \mathbf{k}}$.

\section{Flops at arbitrary field direction}
\label{secarb}

Let us assume now that the external magnetic field
\begin{equation}
\label{mag1}
  \mathbf{h} = h (\sin{t}\cos{f},\sin{t}\sin{f},\cos{t})
\end{equation}
is directed arbitrary. For definiteness, we consider the system with the biaxial anisotropy \eqref{ham1}. An extension to the system with dipolar forces can be made straightforwardly as in Sec.~\ref{Theor2}.
%In Sec.~\ref{Theor1} it was shown that the system response to the in-plane and perpendicular magnetic field is different (see, e.g., Eqs.~\eqref{encorr2} and~\eqref{eclass3}). We also made a conclusion that it is sufficient to treat anisotropy energy on the mean-field level. We use the two statements above for calculations in this Section.
Let us characterize the spiral plane by the vector normal to it
\begin{equation}\label{norm1}
  \mathbf{n}(\theta,\varphi) = (\sin{\theta}\cos{\varphi},\sin{\theta}\sin{\varphi},\cos{\theta}).
\end{equation}
It is convenient to introduce two components of the magnetic field: perpendicular to the spiral plane ${\bf h}_n$ and the in-plane component ${\bf h}_\tau $ whose values read as
\begin{eqnarray}
  h_n &=& h \left[ \sin{\theta} \sin{t} \cos{(\varphi-f)} + \cos{\theta} \cos{t} \right], \\
  h_\tau &=& \sqrt{h^2-h^2_n}.
\end{eqnarray}
In terms of these quantities, the system energy has the form
\begin{equation}
\label{en2}
 \frac{{\cal E}(\theta,\varphi)}{NS^2} \simeq -\frac{E (\cos^2{\varphi} + \cos^2{\theta} \sin^2{\varphi} ) + D \sin^2{\theta} }{2} - \frac{h^2_n}{2 \tilde{J}S^2},
\end{equation}
where the angle-independent term $-J_\mathbf{k}/2 - h^2/2S^2(2 J_\mathbf{k}- J_{\bf 0} -  J_{2 \mathbf{k}})$ is omitted and $\tilde J$ is given by Eq.~\eqref{jtild}.

We analyze now the stability of the spiral planes with respect to small variations in $\theta$ and $\varphi$ using Eq.~\eqref{en2}. Let us start with spin rotation in $yz$ plane (i.e., $\theta=\pi/2, \, \varphi=0$). In particular, energy \eqref{en2} is minimal in this case at $h=0$. Let us discuss the stability of such spin texture at finite magnetic field by considering angle variations of the form
\begin{equation}
\label{var1}
  \theta=\frac{\pi}{2} - \delta \theta, \quad \varphi=\delta \varphi.
\end{equation}
The energy variation reads as
\begin{eqnarray} \nonumber
  \frac{\delta{\cal E}(\theta,\varphi)}{NS^2} &=& \frac{E (\delta \varphi)^2 + D ( \delta \theta)^2 }{2}\\
	&&-\frac{h^2}{\tilde{J}S^2}(\delta \theta \cos{t}  + \delta \varphi \sin{t} \sin{f} )\sin{t} \cos{f}\nonumber\\
	&&-\frac{h^2}{2\tilde{J}S^2} \bigl[ (\delta \varphi)^2 \sin^2{t}(\sin^2{f}-\cos^2{f}) \label{var2}\\ &&+2 \delta \theta \delta \varphi \cos{t} \sin{t} \sin{f} \nonumber\\  &&+ (\delta \theta)^2 (\cos^2{t} - \sin^2{t} \cos^2{f})  \bigr] \nonumber
\end{eqnarray}
Notice that there are field-dependent terms in Eq.~\eqref{var2} linear in $\delta \theta$ and $\delta \varphi$. They vanish if magnetic field lies in $yz$ plane (i.e., at $f=\pi/2$) and if $\bf h$ is parallel to $x$ axis (i.e., at $t=0$). In other cases, linear terms lead only to a continuous rotation of the spiral plane by the external magnetic field ($\mathbf{n}(\theta,\varphi)$ rotates towards the magnetic field direction).

No spiral plane flops can happen also if the the magnetic field is oriented along $x$ axis because $h$-dependent terms in Eq.~\eqref{var2} read in this case as
\begin{equation}
\label{var4}
  \frac{h^2}{2 \tilde{J}S^2} \left[ (\delta \varphi)^2 +  ( \delta \theta)^2 \right]
\end{equation}
that results in a stable energy minimum for the spin texture in $yz$ plane.

If $\bf h$ lies in $yz$-plane (i.e., at $f= \pi/2$), we have for $h$-dependent terms in Eq.~\eqref{var2}
\begin{equation}
\label{var5}
-\frac{h^2}{2 \tilde{J}S^2}  \left[ (\delta \varphi)^2 \sin^2{t} + 2 \delta\theta \delta\varphi \cos{t} \sin{t} +  ( \delta \theta)^2 \cos^2{t} \right].
\end{equation}
The energy minimum at $\theta=\pi/2$ and $\varphi=0$ is stable until $\delta{\cal E}(\theta,\varphi)$ remains a positively defined quadratic form, i.e., if the following inequality holds:
\begin{equation}
\label{flop1}
  ED - \frac{h^2}{\tilde{J}S^2} (E \cos^2{t} + D \sin^2{t}) > 0.
\end{equation}
The field value at which the spiral plane flop takes place can be found from Eq.~\eqref{flop1} with the result
\begin{equation}
\label{flop2}
  h_{flop} = S\sqrt{ \tilde{J} \frac{ED}{E \cos^2{t} + D \sin^2{t}}}
\end{equation}
which is a generalization of Eq.~\eqref{hc12} for arbitrary $t$. The generalization of Eqs.~\eqref{hc14} and \eqref{hc15} has the form
\begin{equation}
\label{flop2dip}
  h_{flop} = S\sqrt{ \tilde{J} \frac{\left[\lambda_1(\mathbf{k})-\lambda_2(\mathbf{k})\right]\left[\lambda_1(\mathbf{k})-\lambda_3(\mathbf{k})\right]}{\left[\lambda_1(\mathbf{k})-\lambda_2(\mathbf{k})\right] \cos^2{t} + \left[\lambda_1(\mathbf{k})-\lambda_3(\mathbf{k})\right] \sin^2{t}}}.
\end{equation}
The generalization of Eq.~\eqref{hc13} reads as
\begin{equation}
\label{hc132}
  h_{flop} = \sqrt{2 S h_s \frac{ED}{E \cos^2{t} + D \sin^2{t}}}
\end{equation}

Let us discuss now the orientation of the spiral plane after the flop when $\bf h$ lies in $yz$ plane. We have carried out an analysis of the stability of the configuration with $\theta=t$ and $\varphi=f$ similar to that performed above. We have found that the anisotropy provides terms in the energy linear in angles variations if the field is not directed along $y$ or $z$ axes. Thus, we make a conclusion that if the external magnetic field is in $yz$-plane but $\theta \neq 0$ or $\pi/2$, $\mathbf{n}$ is not parallel to $\bf h$ after the flop and it smoothly rotates towards $\mathbf{h}$ upon further field increasing.

\section{Possible applications}
\label{appsec}

We discuss in this section application of the theory proposed above to particular spiral materials.

%MnWO$_4$ undergoes cascade of field-induced phase transitions at low temperatures. \cite{mnw1} It has a very complicated phase diagram in $H$--$T$ plane. \cite{Ehrenberg1997} It is shown in Ref.~\cite{Taniguchi2006} that the elliptical spiral phase is ferroelectric at intermediate temperatures and that the electric polarization can be accounted for using the inverse Dzyaloshinsky-Moria (DM) mechanism. Moreover, the electric polarization can be flopped to another direction by means of external magnetic field. The in-plane magnetic field leads to reorientation of the spin rotation plane in the elliptical phase. \cite{Urcelay2014} In the recent numerical study~\cite{zh} it was demonstrated that the magnetic phase diagram of MnWO$_4$ can be obtained using just frustrated exchange interaction and biaxial anisotropy, the observed complication of phase diagram near the flop field $h_{flop}$~\cite{Urcelay2014} is due to the close commensurate modulation vector. We adopt dimensionless parameters (we consider them as measured in K) from Ref.~\cite{zh} and using our approach we obtain $h_c \approx 24~\text{K}$ and $h_{flop} \approx 7.25~\text{K} $ for the saturation and the flop fields, respectively. These values are in a very good agreement with the numerics of Ref.~\cite{zh}.

Co-doped MnWO$_4$ with the dopant concentration $0.05$ is thoroughly investigated experimentally in Ref.~\cite{Urcelay2017}. Mn$_{0.95}$Co$_{0.05}$WO$_4$, in contrast to pure MnWO$_4$~\cite{mnw1}, is in a multiferroic cycloidal phase at small $T$. Application of in-plane magnetic field leads to a spontaneous flop of the spin rotation plane perpendicular to the field at
$h=h_{flop}\approx10\mbox{ \rm T}\ll h_s\approx 60~\text{T}$. \cite{Urcelay2017}
If $\bf h$ is directed along the hard axis, the spin rotation plane stays intact. This picture is very similar to that we obtain above theoretically. The difference is that for $\bf h$ directed along the medium axis the flop is replaced by a rather rapid but continuous rotation of the spiral plane in a field interval of about 4~T. The latter may be attributed to local anisotropy of Co ions and requires more careful consideration. Since Mn$^{2+}$ ions are in spherically symmetric state with $L=0$ and $S=5/2$, it is expected that the anisotropy of the spin-orbit origin is strongly suppressed and the main anisotropic interaction in the system is the dipolar one. We have calculated eigenvalues of the dipolar tensor ${\cal D}^{\alpha \beta}_\mathbf{q}$ for pure MnWO$_4$ and substitute them to Eqs.~\eqref{hc14} and \eqref{hc15} for $h_{flop}$ estimation in Mn$_{0.95}$Co$_{0.05}$WO$_4$. Values of $J_{\bf 0}$, $J_{\bf k}$, and $J_{2\bf k}$ arisen in Eqs.~\eqref{hc14} and \eqref{hc15} have been calculated using exchange coupling constants found from fitting of neutron experimental data in Ref.~\cite{Xiao2016}. For magnetic field along easy axis, we find $h_{flop} = 8~\text{T}$ while experimentally observed~\cite{Urcelay2017} value is $\approx 10~\text{T}$. For magnetic field directed along medium axis, we obtain $h_{flop} = 6.5~\text{T}$ which lies in the middle of the field interval, where the continuous rotation of the spiral plane is observed experimentally~\cite{Urcelay2017}. Notice also that $h_{flop}$ found using Eq.~\eqref{jta} via experimentally obtained $h_s$ is only 20\% smaller than that obtained above although $J_{\bf 0}$ is 1.5 times as large as $J_{2\bf k}$.

EuNiGe$_3$ is a helimagnet with equally possible spiral vectors $\mathbf{k}=(\frac14,\delta,0)$, $(\frac14,-\delta,0)$, and $(\delta,\frac14,0)$ allowed by the tetragonal symmetry, where $\delta=0.05$. \cite{Gukasov2016} Magneto-dipolar interaction is expected to be very important in this material because exchange constants are rather small and Eu$^{2+}$ ions are in a spherically symmetric state with $L=0$ and $S=7/2$. \cite{Gukasov2016} It can be shown~\cite{UtesovEu} that dipolar forces make the spiral plane to be perpendicular to $\mathbf{k}$ in agreement with experimental observations. It is believed that a small Dzyaloshinskii-Moriya interaction is responsible for the finite $\delta$. \cite{Maurya2014} Magnetic field directed along $a$ and $b$ tetragonal axes results in the spiral plane flop accompanied with changing $\bf k$ by another equivalent spiral wave vector. \cite{Gukasov2016} Then, the theory presented above should be modified to describe such flops (as it is done in Ref.~\cite{syrom} for a collinear antiferromagnet). However, $\bf k$ does not change significantly during the flop if $\bf h$ is parallel to $c$ axis and our theory can work in this case. Calculations show that $\lambda_a(\mathbf{k})-\lambda_c(\mathbf{k}) =0.135~\text{K}$ in Eq.~\eqref{hc14}. To estimate $\tilde J$ given by Eq.~\eqref{jtild} and appearing in Eq.~\eqref{hc14}, we assume that $J_{2\bf k}\approx J_{\bf 0}$ in which case $\tilde J$ is related to $h_s$ (see discussion after Eq.~\eqref{jta}). It was found experimentally that the saturation field $h_s\approx 6~\text{T}$ in EuNiGe$_3$. \cite{Gukasov2016} As a result, we obtain $h_{flop}=2.05~\text{T}$ which matches excellently the experimentally observed value $\approx 2~\text{T}$. \cite{Gukasov2016}

Spiral plane flops have been reported recently also in many others spiral magnets many of which are multiferroics: LiCu$_2$O$_2$, \cite{Park2007,Bush2012,Bush2018} NaCu$_2$O$_2$, \cite{Sadykov2014} CuCrO$_2$, \cite{Seki2008,Kimura2009,Soda2010} CuCl$_2$, \cite{Seki2010} LiCuVO$_4$, \cite{Naito2007,Buttgen2007,Schrettle2008} and KCu$_3$As$_2$O$_7$(OD)$_3$ \cite{Nilsen2014} to mention just a few. In all of them the anisotropy of spin-orbit origin is expected to overcome significantly the dipolar forces. On the other hand, values of anisotropy have not been determined yet in these compounds so that we cannot check our theory in these cases.

\section{Summary and conclusion}
\label{conc}

To conclude, we present a theory of field-induced flops of plane in which spins rotates in frustrated Heisenberg helimagnets with small anisotropic interactions, biaxial anisotropy and dipolar forces. We find that flops occur upon the field increasing if the field lies in the spiral plane stabilized at $h=0$. The spiral plane becomes perpendicular to the field after the flop (see Fig.~\ref{Fig1}). The critical fields $h_{flop}$ are given by Eqs.~\eqref{flop2} and \eqref{flop2dip} for biaxial anisotropy and dipolar interaction, respectively. In the case of biaxial anisotropy, if $J_{\bf 0}\approx J_{2\bf k}$, where $\bf k$ is the helix vector, $h_{flop}$ is expressed via the saturation field $h_s$ (see Eq.~\eqref{hc132}) that opens a simple way to determine the anisotropy value if $h_{flop}$ and $h_s$ are known. Notice also that if the field is directed along the easy axis Eq.~\eqref{hc132} is identical to that for the spin-flop field in collinear axial magnets. In contrast to the spin flop in collinear magnets, where the flop takes place only at a very narrow interval of the field directions along the easy axis, \cite{bogdan} flops of the spiral plane happens at any orientation of the field in the spiral plane.

\begin{acknowledgments}

We thank S.V.\ Maleyev for stimulating discussion. The reported study was funded by RFBR according to the research project 18-02-00706.

\end{acknowledgments}

\appendix

\bibliography{TAFbib}

\end{document}